 \documentclass[usenatbib,onecolumn]{mnras}

\usepackage{wrapfig}
\usepackage{multicol}

\usepackage{graphicx}
\usepackage{natbib}
\usepackage{pdflscape} 
\usepackage{amssymb}
\usepackage{ulem} 
\usepackage{color} 
\usepackage{amsmath} 

\newcommand{\be}{\begin{equation}}
\newcommand{\ee}{\end{equation}}
\newcommand{\nn}{\mbox{} \nonumber \\ \mbox{} }
\newcommand{\ba}{\begin{eqnarray}}
\newcommand{\ea}{\end{eqnarray}}

\newcommand{\Alfven}{ Alfv\'{e}n }

\newcommand{\curl}{{\rm curl\, }}

\newcommand{\B}{{\bf B}}

\newcommand\eg{{\it{{e.g., }}}}

\newcommand{\Bf}{{magnetic field}}

\newcommand{\NS}{neutron star}

\newcommand{\ms}{magnetosphere}
\newcommand{\mss}{magnetospheres}

\def\1023{PSR J1023+0038}

\def\m28{IGR J18245--2452}

\begin{document}

\title{Centrifugal barriers in  magnetospheric accretion}

\author{Maxim Lyutikov\\
Department of Physics and Astronomy, Purdue University,  525 Northwestern Avenue, West Lafayette, IN, USA }

\maketitle

\begin{abstract}
We reconsider  the dynamics of  accretion flows onto magnetized central star.  For    dipolar  magnetically aligned  case, the centrifugal barrier   is  at 
$R_{cb} = (2/3)^{1/3} R_c = 0.87 R_c$, where $R_c= ( G M/\Omega^2)^{1/3}$ is the corotation radius.  
 For oblique dipole   direct accretion from the corotation radius  $R_c$ is possible only for  magnetic obliquity  satisfying  $\tan \theta_\mu \geq 1/( 2 \sqrt{3}) $ ($\theta_\mu \geq  16.1^\circ $).  The accretion   proceeds  in a form of funnel flows - along two streams centered on the $\mu-\Omega$ plane, with azimuthal opening  angle $ \cos (\Delta  \phi)  =  { \cot^ 2 {\theta_\mu} }/{12} $.  For the \ms\ distorted by the diamagnetic disk, the centrifugal barrier can be at as small radius as $R_{cb}= 0.719 R_c$ for the fully confined dipole, extending out to $R_{cb} \sim  R_c$ for the magnetically balanced case.   Type-II X-ray bursts in accreting {\NS}s may be  mediated by the centrifugal barrier; this requires nearly aligned configuration. Centrifugally-barriered material trapped in the \ms\ may lead to periodic obscuration (``dips'') in the light curve of the host star,  \eg as observed  in accreting young stellar objects and X-ray binaries.
\end{abstract}

\keywords{Physical data and processes: accretion, accretion discs; magnetic fields}

\section{Introduction}

Accretion onto a magnetized star is a classical  problem in  astrophysics. This includes accretion onto {\NS}s,  \citep{1975JETP...41...52B,1979ApJ...232..259G,1977ApJ...217..578G,1997ApJS..113..367B,2022arXiv220414185M},  accretion onto WDs - Intermediate Polars \citep[][]{2015SSRv..191..111F},   and in 
young protostars \citep{1994ApJ...429..781S,2007ARA&A..45..565M,2009A&A...508.1117Z,2013A&A...550A..99Z}.

In  an accepted  picture \citep{1973ApJ...184..271L,1977ApJ...217..578G,1992ans..book.....L}, steady  state accretion disks tends to have the inner edge of the accretion disk near the corotation radius. The disk material first moves inward through the disk \citep{1973A&A....24..337S}, and later on couples, through some kind of dissipative processes \citep{1976ApJ...207..914A} to the \ms\ and flows onto the central star.

Many models agree on the fact that for accretion to occur the truncation radius  $R_{in}$
of the  disk is  located inside the corotation radius \citep[\eg][]{1997ApJ...489..890M}. For oblique case, accretion takes a form of 
funnel flows \citep{2002ApJ...578..420R,2008A&A...478..155B}.

 In this paper we consider effects of gravitational attraction and centrifugal barrier. As the plasma inside the inner edge of the disk is expected to be magnetically dominated, it can flow mostly along \Bf\ lines. We consider several cases (pure  dipole and/or \ms\ distorted by the currents in the disk \citep[see][for a somewhat related treatment]{2020MNRAS.496...13A}.  
Presence of the accretion barrier has been discussed  in a number of works \citep{1977SvAL....3..138S,1992ans..book.....L,1993ApJ...402..593S,2012MNRAS.420..416D}. It was typically  assumed that the barrier is located outside of the corotation. We demonstrate that it is slightly inside for the aligned case.

\section{Dynamics of magnetospheric  accretion}

\subsection{The basics:  pure dipole, aligned case $\theta _{\mu}=0$} 

In the present work we consider the dynamics of the flows from the disk onto the central star. As the inflowing matters enters the \ms\ of the host, it's dynamics quickly becomes controlled   by the ever-increasing \Bf.  The matter mostly flows along the rotating  fields lines. Two forces are at play: gravitational  attraction towards the star, and the centrifugal repulsion of the material following the field lines. The actions of the  gravitational and centrifugal forces, combined with restricted motion  along the \Bf\ lines create complicated dynamical system.

There is gravitational attraction acceleration
\be
{\bf F} _g =-   \frac{ G M}{r^2} {\bf e}_r
\ee
($ {\bf e}_r$ is  a spherical radial unit vector).

For matter bound to the    \Bf\ lines rotating with the star at frequency $\Omega$  there is also  a centrifugal acceleration  
\be
{\bf F} _c = r \Omega^2 {\bf e}_\varpi
\ee
($ {\bf e}_\varpi $  is  a cylindrical  radial unit vector). The total force acting on a particles  is ${\bf F} _{tot} = {\bf F} _g+ {\bf F} _c$.

For a particle trying  to fall onto the  magnetized  central star along the dipolar \Bf,  the important parameter is the projection of the force  ${\bf F} _{tot}$ along the local \Bf. For some configurations the resulting force will try to pull the particles inward (hence accretion), for some cases the centrifugal repulsion (along the \Bf) will dominates  -  the centrifugal barrier.

In what follows we assumes that the disk plane is perpendicular to the spin of the central star - the spin is along $z$ axis, the disk is in the $x-y$ plane.

Let's consider the aligned case, when the magnetic moment is aligned with the spin. In the (arbitrary) $x-z$ plane the unit \Bf\ is 
\be
{\bf b}= \left\{\frac{3 \sqrt{2} \sin \theta  \cos \theta }{5+ \sqrt{3 \cos (2 \theta )}},\frac{ 1+ 3
   \cos (2 \theta )}{\sqrt{10+6 \cos (2 \theta )}}\right\}
   \label{b} 
   \ee
(this is unit \Bf\ vector in Cartesian coordinates $\{x,z\}$, expressed as function of polar angle $\theta$). 

We find for the force parallel to the \Bf:   
   \ba &&
   F_\parallel  = ({\bf F} _{tot}  \cdot {\bf b}  )  \propto \left(3 \tilde{r}^3 \cos (2 \theta )-3 \tilde{r}^3+4\right) \cos \theta  
   \nn &&
   \tilde{r} = \frac{r}{R_c}
   \nn &&
   R_c = \left( {\frac{G M}{\Omega ^2}} \right)^{1/3}
   \ea
     
    \begin{figure}
\centering
\includegraphics[width=.99\linewidth]{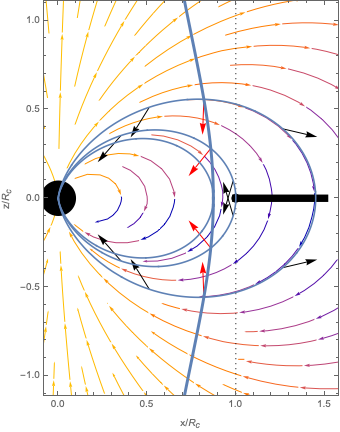} 
\caption{Geometry of forces for aligned rotator.  Distances are normalized to the corotation radius $R_c$. Arrows (normalized to a fixed length) indicate the sum of gravitational and centrifugal forces. The thick line, Eq. (\protect\ref{null}), indicates where the total force is perpendicular to the local \Bf. To the left of that line particles are pulled towards the star, to the right they are pushed away by the centrifugal forces. There is a gap that would prevent accretion from the disk ending at the corotation radius. }
\label{aligned}
\end{figure}

Condition $ F_\parallel  =0$ gives the location of the null points - where the force of gravitational  attraction and centrifugal repulsion, as projected along the \Bf\ line, balance out: 
\be
 \tilde{r}  = \frac{  (2/3)^{1/3}} {\sin ^{2/3} \theta}
 \label{null}
 \ee
 see Fig. \ref{aligned}; the case $\tilde{r} =1$, $\theta =\pi/2$ is just a special singular point. The scaling  $\sin ^{-2/3} \theta$ is known \citep{1992ans..book.....L}.
 
 Importantly, the centrifugal barrier is located {\it inside}  the corotation radius,  extending from $  (2/3)^{1/3}R_c $ to $R_c$. Within the barrier   particle leaving the disk would be pushed back by the dominant centrifugal force.  (The field lines that extend to $R_c$ crosses the null line at 
$
\sin \theta = (2/3)^{1/8}
$).
Taken at face value: for an aligned rotator accretion from  the co-rotation radius is impossible - the disk should extend to the inner radius  $R_{in}$ less than the radius of the centrifugal barrier $ R_{cb} $,  $R_{in} \leq R_{cb} \leq R_c$.

  \subsection{Oblique  dipole $\theta _{\mu} \neq 0$} 
  
\begin{figure} 
\centering
\includegraphics[width=.99\linewidth]{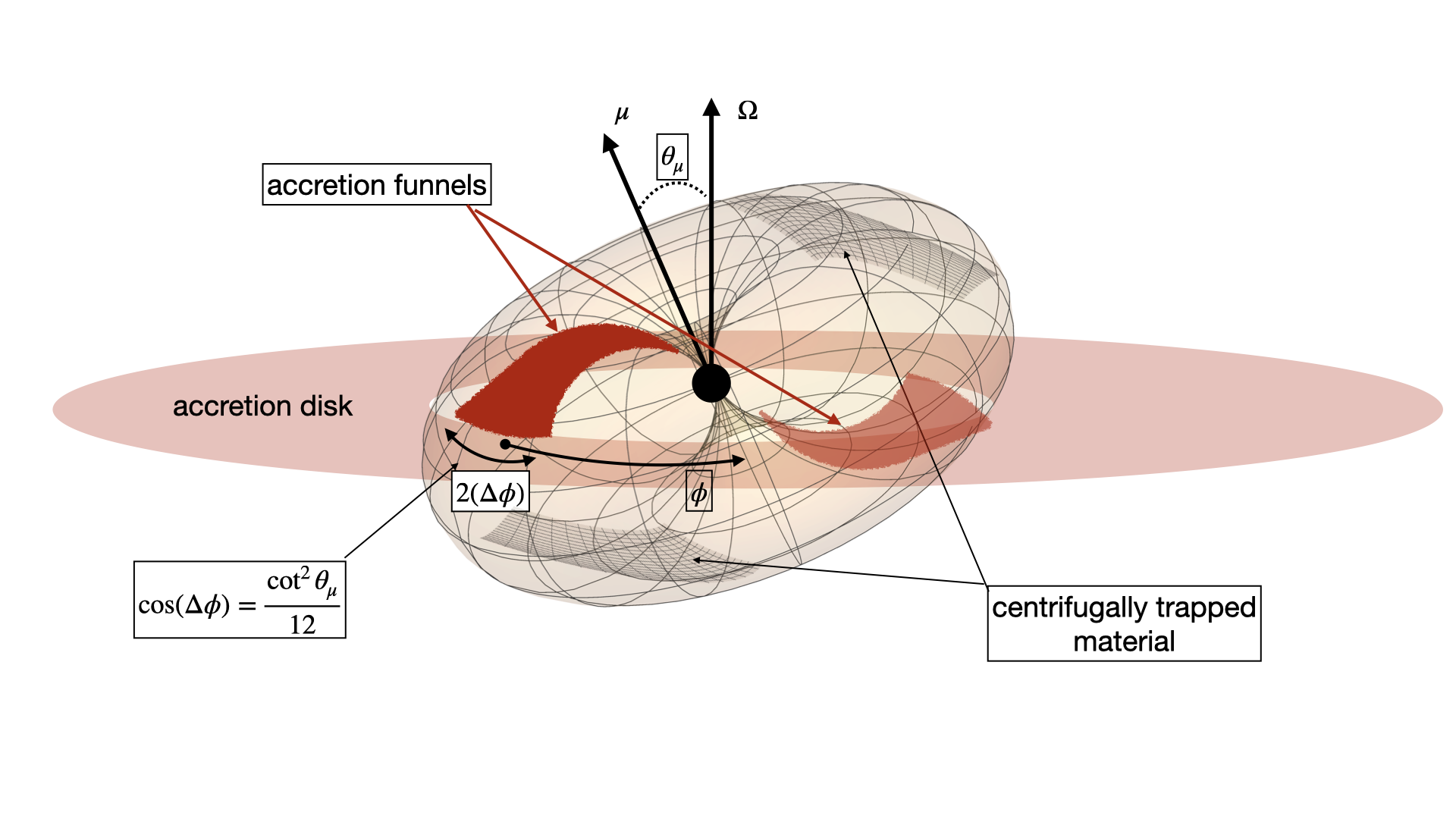}
\caption{ Cartoon of the accretion  flow onto magnetized central star. At every moment there are two accretion streams, flowing  mostly in the $\mu-\Omega$ plane.  At  the opposite part of each  magnetic hemisphere centrifugally-barriered  material accumulates. Angle $\phi$ is measured from the $\mu-\Omega$ plane.
}
\label{cartoon}
\end{figure}

  Next we extend the results to the oblique case. Using relations for the \Bf\ of an oblique dipole one can find location of the points where the total force is perpendicular to the \Bf\  as a function of azimuthal angle $\phi$:
   \ba &&
 \tilde{r}  = \left(  \frac{4 \cos \Theta }{ \cos \Theta (1  -3 \cos (2 \theta ) )  +2 \cos \Theta  \cos
  \theta _{\mu }}\right)^{1/3} 
   \nn &&
    \cos \Theta  = \sin \theta  \cos \phi \sin\theta _{\mu }+\cos \theta  \cos
  \theta _{\mu } 
     \ea 
     (the dipole moment is in the plane $\phi=0$, see Fig. \ref{cartoon}).

\begin{figure}
\centering
\includegraphics[width=.9\linewidth]{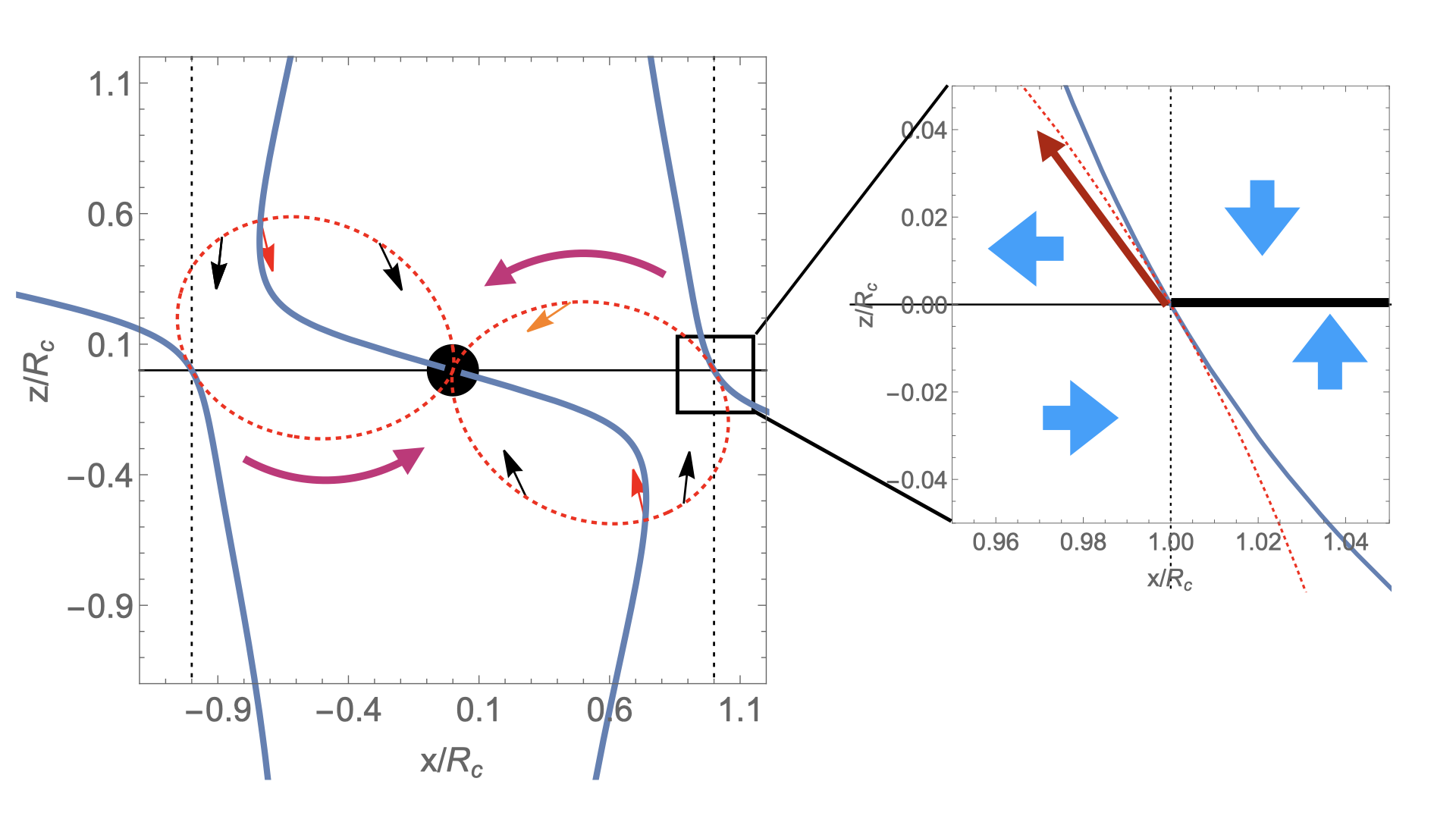} \\
\includegraphics[width=.9\linewidth]{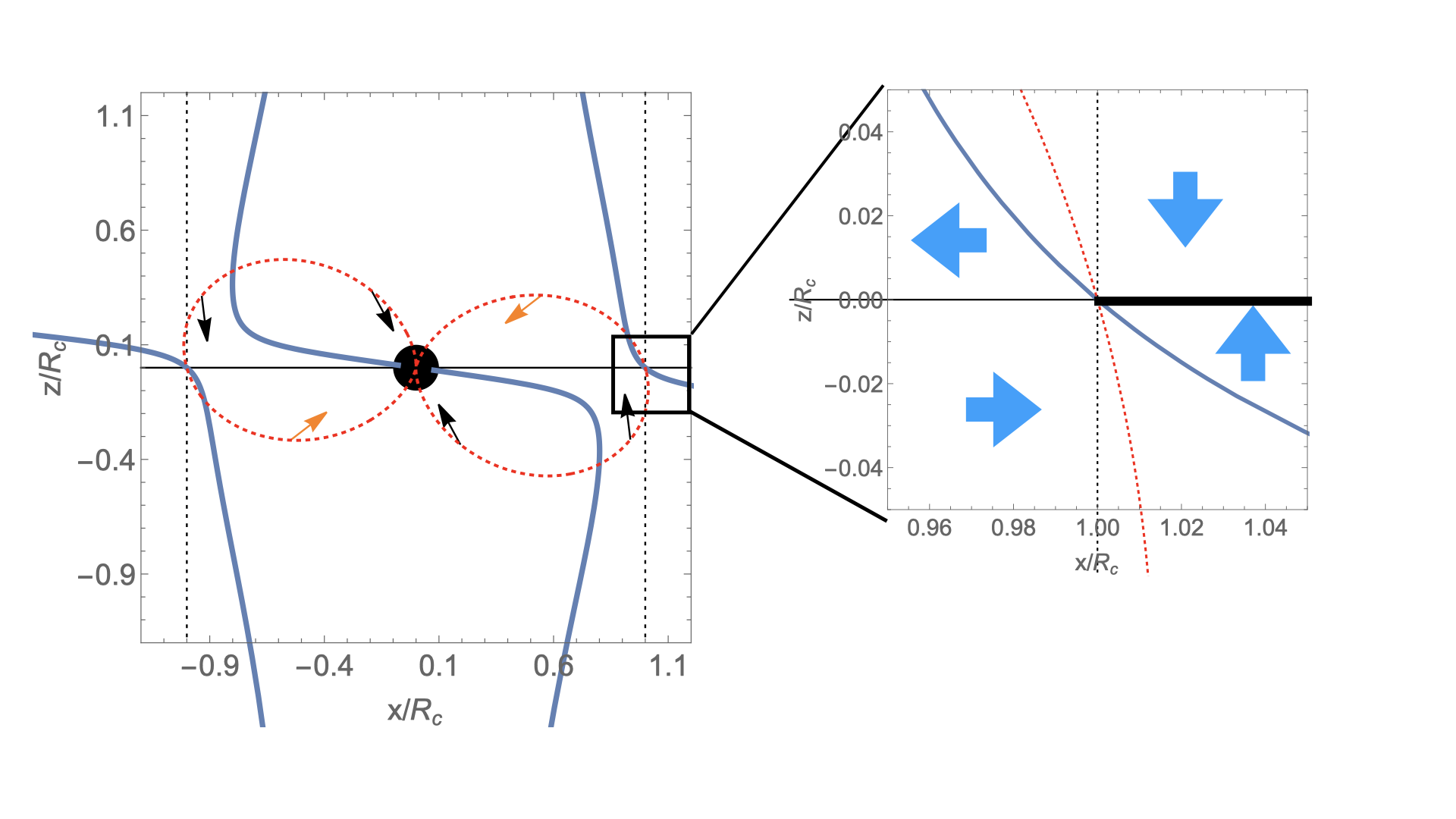} 
\caption{Critical case and sub-critical case  ($\theta_d =\theta_d^{(crit)}/2$). Blue arrows indicate general direction of force on particles outside the disk.  Particles may from from the inner edge of the disk along  the \Bf\ line (dashed red).  For the inflow to occur the dashed red line should be to the left of the null line. In the subcritical case the total gravitational plus centrifugal force prevents such a low. Thick maroon arrows in the critical case indicate possible direction of the flow.}
\label{oblique}
\end{figure}

  For the flow in the $\mu-\Omega$ plane the  null line is now at 
\be
 \tilde{r}  = 2^{2/3} \frac{ \cos ^{1/3} (\theta-\theta_\mu)  }{ \sin^{1/3} \theta  \left( 3 \sin (2 \theta-\theta_\mu) +\sin \theta_\mu\right)^{1/3} } , 
  \label{null1}
 \ee
 Fig. \ref{oblique}. 

For any $\theta_\mu \neq 0$ there are field lines that start in the equatorial plane just inside the corotation, and connect to the star always with inward-directed projection of the total force. If oblique disk extends to $\leq R_c$ the accretion can be continuous.  For the field line connecting to the corotation radius
there is a critical obliquity angle $\theta_\mu ^{(crit)}$, 
\be
\tan \theta_\mu ^{(crit)}= \frac{1}{2 \sqrt{3}}
\ee
For   $\theta_\mu \leq  \theta_\mu ^{(crit)}$ there exists a ``buffer'' -  a region at $r< R_c$ where the centrifugal force dominates over the gravitation attraction (hence, formally preventing accretion). For $\theta_\mu \geq  \theta_\mu ^{(crit)}$ a channel for accretion opens up: but only in one hemisphere, for each corotation point.

\begin{figure} 
\centering
\includegraphics[width=.49\linewidth]{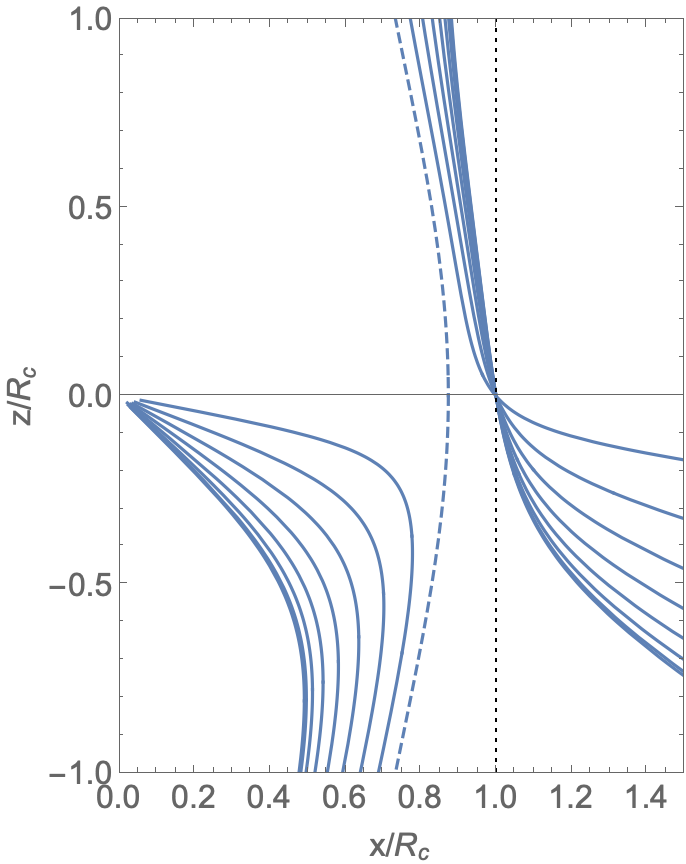}
\includegraphics[width=.49\linewidth]{barrierphi1.png}
\caption{ Location of the centrifugal barrier for different azimuthal angles $\phi =0, \,  \pi/16... \pi/2$ for two magnetic inclination angles $\theta_\mu = \pi/4$ and $\theta_\mu =  \theta_\mu ^{(crit)}$. Dashed lines are for $\phi =\pi/2$. At these points the accretion flow is the same as for the aligned case. This picture  illustrates that different azimuthal angles have different centrifugal barriers -  as  a result, generically, one expects two accretion funnels.
}
\label{barrierphi}
\end{figure}

\begin{figure} 
\centering
\includegraphics[width=.99\linewidth]{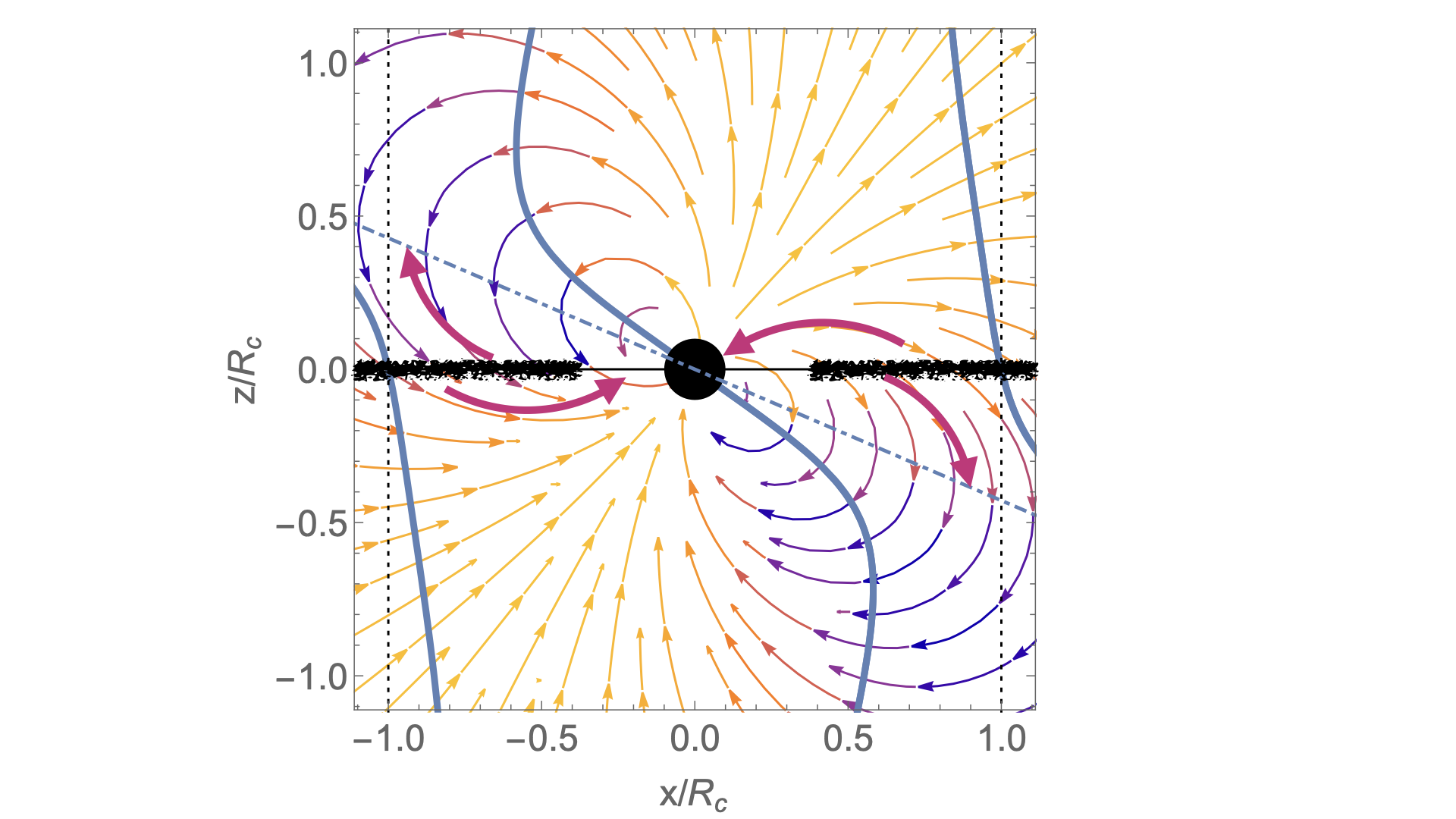}
\caption{Illustration of the magnetospheric structure for $\theta_\mu=\pi/4$, cut along $\phi = \pi/4$ plane.  Solid lines are centrifugal barriers. Dot-dashed lines are the points of furthest extension of a given field lines, Eq. (\protect\ref{furthest}).  Centrifugally-barriered material is expected to accumulate there, see \S \ref{fate}. To supply such material the disk should extend to $ \leq R_c$. Maroon lines indicate the possible matter trajectories.}
\label{thetapi4phipi4}
\end{figure}

   For the general case (oblique dipole, flow starting at arbitrary azimuthal angle),  there is a clear interesting result: flow in the 
 orthogonal plane $\phi =  \pm \pi/2$.    We find that in that plane the relations reduce to the aligned  case (\ref{null}). The centrifugal barrier between $R_{cb}$  and $R_c$ would prevent accretion from the $R_c$ at that location, Fig. \ref{barrierphi}-\ref{thetapi4phipi4}.
       
    For accretion from the corotation radius at arbitrary azimuthal angle $\phi$   the critical  magnetic obliquity angle  becomes
\be
\tan \theta_\mu ^{(crit)}= \frac{1}{2 \sqrt{3 \cos \phi}}
\ee
($\phi=0 $ corresponds to the $\Omega-\mu$ plane). Thus, for a given magnetic obliquity angle $\theta_\mu$ accretion from the corotation radius can proceed only for 
\be
\cos \phi \leq  \frac{ \cot^ 2 {\theta_\mu} }{12}
\ee

 \subsection{Centrifugally-barriered material} 
 \label{fate} 
 
 Material will be centrifugally pushed away along the field lines in the regions where centrifugal force dominates over gravity (as projected along a given field). In the bead-on-wire approximation then the material will be accumulated at the furthest points given by
 \be
 (1-3 \cos (2 \theta )) \cos \phi \sin \theta _{\mu }+3 \sin (2 \theta ) \cos \theta _{\mu } =0
 \label{furthest}
 \ee
 (For given inclination angle $\theta _{\mu }$ and given azimuth $\phi$ (with respect to the $\Omega-\mu$ plane)  this relation determines the location of points of furthest radial extent of a given field lines, Fig. \ref{thetapi4phipi4}-\ref{cartoon}.

  As enough material is accumulated, bead-on-wire approximation will be broken, and material will be  expelled from the \ms.
  Estimating moment of inertial of the trapped plasma as
\ba &&
I_{trapped} \sim \rho R_c^5 \sim \frac{ B_\ast^2 R_\ast^6}{ 8 \pi G M_\ast} 
\nn && 
 \rho (R_c \Omega)^2 \sim \frac{B(R_c)^2}{8 \pi} 
 \ea 
 The value of $I_{trapped}$ is tiny compared with the moment of inertia of the central star, hence no effect on the timing of the central star is expected.


\section{Centrifugal barrier in magnetospheres distorted by the disk} 
\label{Dia}
 
 \cite{1980A&A....86..192A} and 
 Aly   (priv. comm., unpublished) \citep[see also][]{1980Ap&SS..71..195R,1985IAUS..107..217A,1990A&A...227..473A} found a sequence of  solution for a dipole plus diamagnetic disk. Solutions extend from configurations with all fields closed,
 \ba &&
 X= \sqrt{2} a \cos (\theta ) \sqrt{\frac{1}{-a^2+{r^2}/{Y}+r^2}}  
 \nn &&
 Y= \frac{r^2}{\sqrt{4 a^2 r^2 \cos ^2(\theta )+\left(r^2-a^2\right)^2}} 
 \nn &&
 B_r = \left(\frac{1-Y}{X}+\arctan (X)\right) \frac{4 \cos (\theta ) }{\pi  r^3} \mu 
 \nn &&
 B_\theta = \left(\frac{\cot ^2(\theta ) \left(\frac{X^2 Y \left(a^2+r^2\right) \sec ^2(\theta
   )}{a^2}-1\right)}{X}+\arctan (X)\right) \frac{2 \sin (\theta ) }{\pi  r^3}\mu,
   \label{Aly1}
   \ea
   to the fully   {\it balanced}  magnetic dipole confined by the disk (there is no magnetic stress at the inner edge):
 \ba  && 
A_{\phi} =\left[  \frac{2}{\pi}  \left( \frac{\tilde{X} }{a^2 \sin  \theta }+\frac{\sin  \theta  \arctan (\tilde{X})}{r^2}-\frac{\cos (\theta
   ) \cot  \theta }{r^2 \tilde{X}} \right)  +  \frac{8  }{2 \pi  a  r \sin  \theta }  \left(1-\frac{r }{a} \tilde{X} \right) \right] \mu 
\nn &&
\tilde{X}= a \cos  \theta  \sqrt{\frac{2}{\sqrt{4 a^2 r^2 \cos ^2(\theta
   )+\left(r^2-a^2\right)^2}-a^2+r^2}}
   \nn && 
\B = \curl {\bf A} 
\nn &&
\mu = B_{\ast} R_\ast^3
\label{BalancedDiskExpressions}
\ea
where $a$ is the inner edge of the disk. At the inner edge of the disk the \Bf\ diverges for the  (\ref{Aly1})  case, and is zero zero  for  (\ref{BalancedDiskExpressions}) (hence there is no force on the disk).
The separatrix between the open and closed field lines is at $\theta_1 = - \pi/6$ (the angle between the separatrix and the vertical axis $z$).

\begin{figure} 
\centering
\includegraphics[width=.99\linewidth]{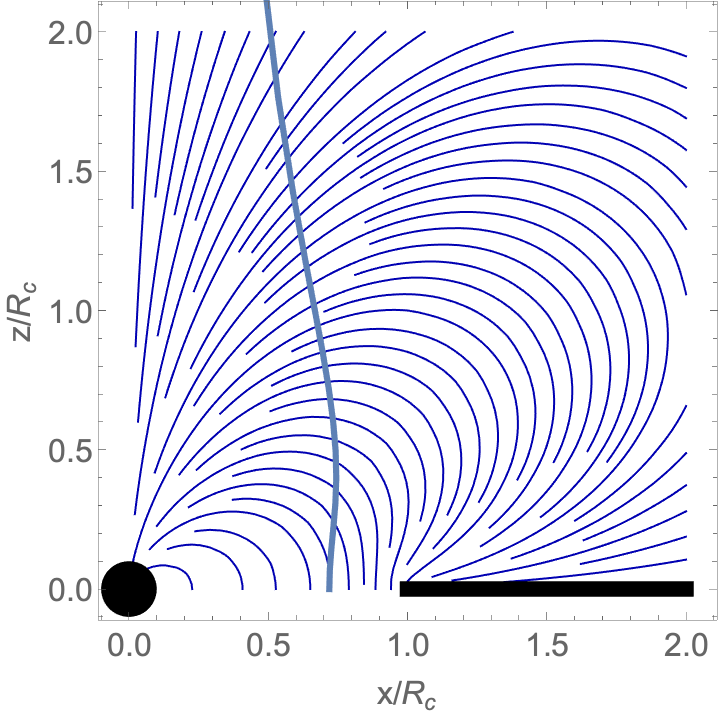}
\caption{Structure of the \Bf\ for  fully confined dipole. Solid line is the centrifugal barrier (it extends down to  $r= 0.719 R_c$  at equator).}
\label{figAly1}
\end{figure}

Assuming that the inner edge is at corotation, $a= R_c$, we find that for the fully confined dipole (\ref{Aly1})  the centrifugal barrier extends down to $r= 0.719 R_c$, Fig.  \ref{figAly1}.

\begin{figure} 
\centering
\includegraphics[width=.49\linewidth]{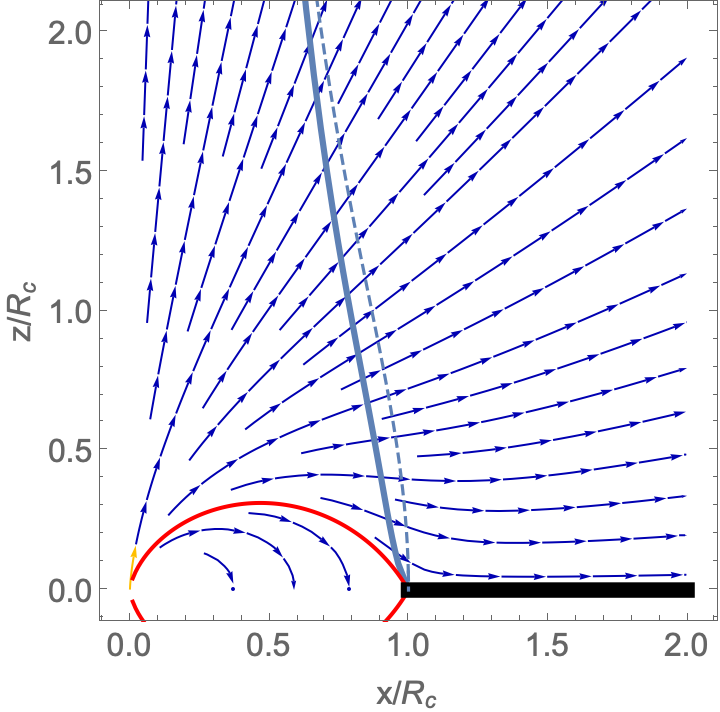}
\includegraphics[width=.49\linewidth]{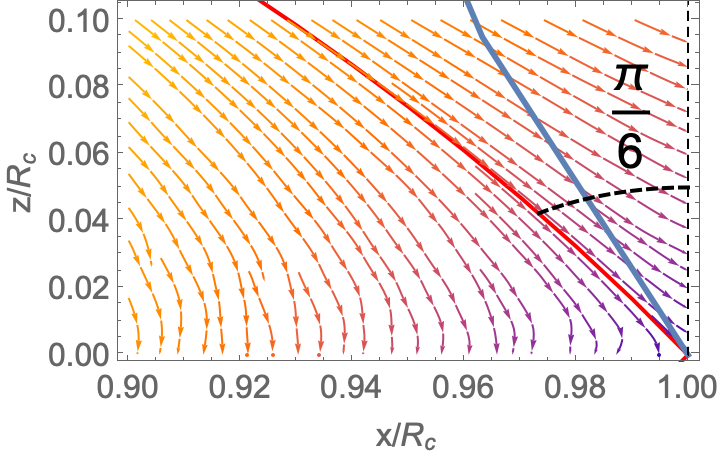}
\caption{Structure of the \Bf\ for a balanced  diamagnetic disk. Red line is the separatrix between open and closed field lines, solid line is the null point (where the combined gravitational and centrifugal force balance out).  Dashed line is approximation for monopolar field (\protect\ref{mono}).  Right panel:  zoomed-in region near the co-rotation point Since the separatrix is always inside the null line, there is no centrifugal barrier for accretion from the corotation point in the case of diamagnetic disk.}
\label{Aly-force}
\end{figure}

For the balanced disk (\ref{BalancedDiskExpressions}),  see Fig. \ref{Aly-force}. Since the field structure becomes nearly monopolar, approximately the null point is at 
\be
\tilde{r} = \frac{1}{\sin^{2/3} \theta}
\label{mono}
\ee

 The centrifugal barrier is at the corotation near equator. Zooming-in, 
near the point $\theta =\pi$, $r=R_C$ the total force  near corotation point  evaluates to 
\be
{\bf F} _{tot} \propto \left\{ 3 \delta r, z  \right\}
\ee
where $ \delta r = r-R_c$. Thus, in this case there is  no centrifugal barrier for the flow from the co-rotation barrier,  Fig. \ref{Aly-force} -  centrifugal barrier is outside the separatrix.  (Since for  the balanced case (\ref{BalancedDiskExpressions})  there is no centrifugal barrier for accretion from the corotation radius,  there is a partially opened configuration when the null line aligns with the separatrix.)

\section{Discussion} 

\subsection{Type-II X-ray bursters}

How does the  accretion proceeds in the (nearly) aligned case resistive (dipolar)  case,  when it is formally prohibited from the corotation radius?  For steady-state accretion the system tends to evolve fairly quickly to the corotation state.  Large accretion rates,  when the \Alfven radius 
\be
r_A = \left(\frac{B_\ast^4 R_\ast^{12} } { 2 G M_\ast \dot{M}} \right) ^{1/7} 
\ee
is smaller than $R_c$,  will spin-up the star, reducing $R_c$ towards $r_A$, while small accretion rate will slow down the star in the propeller regime, increasing $R_c$ towards $r_A$.

Since system evolves towards corotation, but accretion in that case is not allowed, how does accretion occurs? 
Penetration of the accreting material through rotating \ms\ has been considered by a number of authors - theoretically \citep{1976ApJ...207..914A,1993ApJ...402..593S} and numerically \citep[\eg][]{2005ApJ...634.1214L}. The critical issue is the role of  MHD instabilities in  the transition layer. The \Bf, being a ``light fluid" is likely to become ``porous'' via development of tearing and or ballooning/flute instabilities. Rotation adds another complication: shearing of the inner edge of the disk.

  The question is whether instabilities would provide a  nearly constant flow
\citep[similar to the case of  Solar hedgerow prominences][]{2010ApJ...714..618C},  will be dominated by rare major disruptions, or, most likely, a combination thereof. On the one hand, observations of the   type-II X-ray bursts in accreting {\NS}s seem to indicate rare major disruptions:  the fluence in a type II burst is proportional to the waiting time to the next burst \citep{1993SSRv...62..223L}. On the other hand the Rapid Burster occasionally behaves as a normal Low Mass X-ray Binary, not showing  type II bursts - a possible interpretation is that during this state penetration of the material into the \ms\ is dominated by by numerous small-scale instabilities.
The fact that persistent X-ray luminosity (presumably due to continuous plasma penetration and  time-average  type-II X-ray bursts luminosity are of the same order  \citep[but may vary by  a factor of a few ][]{1993SSRv...62..223L} indicates that both precesses  (quasi-steady and catastrophic accretion)  are important.

Our model is not much different from the standard one, that material is accumulated near the inner edge of the disk, and is damped occasionally -  with the subtle difference that matter accumulation occurs not outside the corotation, but outside the centrifugal barrier  at  $R_{cb}$. A clear perdition is that the  type-II X-ray bursters are nearly aligned rotators \citep[\eg][]{2017MNRAS.466L..98V}. This is consistent with the fact that the Rapid Burster MXB 1730-335 does not show pulsations. Breaking of the centrifugal barrier is expected to be nearly axisymmetric in this  case \citep[see also][]{1993ApJ...402..593S,2012MNRAS.421...63R,2015MNRAS.449..268B}.



\subsection{Magnetospheric absorption} 

The material trapped in the \ms\ may cause quasi-eclipses of the parent star, and can shadow the disk.
First, disk shadowing is observed in some protoplanetary disks  \cite[\eg][]{2018ApJ...868L...3M}. In the present model there are four regions of enhanced density with the \ms: 
 two  accretion streams, plus two regions of trapped centrifugally-barriered material. Both can absorb/scatter radiation from the central star. But as they are symmetric, we  expect two shadows  on the outer disk. Two shadows per period are indeed inferred.

 Periodic intensity dips  are  observed \eg in  some T Tauri stars  \citep[``dippers'',][]{1999A&A...349..619B} and X-Ray binaries  \citep{1986ApJ...308..199P}. Both the streams and trapped material can lead to absorption of radiation from the central star, potentially producing two dips per period.  Also, in the critical case, Fig. \ref{oblique} top row, material from the edge of the disk can both be accreted (top hemisphere) and ejected (bottom hemisphere). This can potentially create four streams, two in each hemisphere. 
 
 The  centrifugally-barriered material  will be  periodically centrifugally expelled. No associated timing signal in the rotational properties of the central star is expected.

\section{Conclusion}

We reconsidered the  basic properties of accretion onto a magnetized central star using a number of idealized models for the \ms: pure dipole and dipole plus diamagnetic disk.
First,  for the pure dipolar field, we find that for small magnetic inclination angles accretion from the disk extending down to just the corotation radius is impossible - the inner edge of the disk should be at $R_{cb} = (2/3)^{1/3} R_c = 0.87 R_c$. For higher magnetic obliquity angles accretion proceeds along two counter-aligned streams, the funnel flows. The width of the   funnel flows increases with magnetic obliquity, but they remain separated. If  the inner edge of the disk is at radii smaller that $ (2/3)^{1/3} R_c$ then accretion can proceed at every azimuth point, forming two accretion curtains each spanning $\geq 180^\circ$ near each pole.

Second, we consider \mss\ distorted by the disk - from the complete confinement of the field by the disk, to balanced configurations that produce no force on the disk. 
In these aligned cases the centrifugal barrier can be as deep as $0.71 R_c$ (for the  confined dipole), 

All the models we considered (pure dipole, purely confined dipole, and magnetically balanced disk) are surely mathematical idealization, but they are likely to  ``bracket'' the more realistic cases. Pure dipole is the extreme resistive case (when the disk does not affect the dipolar magnetospheric structure);  purely confined dipole is the extreme ideal case, when the  confined field cannot break out through the disk; magnetically balanced disk assumes that the confined \Bf\ was able to break out and relax to balanced configuration.



 \section{Acknowledgements}
 I would like to thank Pavel Abolmasov, Juri Poutanen, Sergey Tsygankov and  Yanqin Wu for discussions. Important analytical solution (\ref{BalancedDiskExpressions}) is by Jean Jacques Aly. 
 This work had been supported by 
NASA grants  80NSSC18K064,  80NSSC17K0757 and 80NSSC20K0910,   NSF grants 1903332 and  1908590.

\section{Data availability}
The data underlying this article will be shared on reasonable request to the corresponding author.

\bibliographystyle{apj}
 \bibliography{/Users/maxim/Dropbox/Research/BibTex}

\end{document}